**Phase formation, phonon behavior, and magnetic properties of novel ferromagnetic La$_3$BAlMnO$_9$ (B = Co or Ni) triple perovskites**

**M.P. Singh, K.D. Truong, S. Jandl, and P. Fournier**

Département de Physique and Regroupement québécois sur les matériaux de pointe, Université de Sherbrooke, Sherbrooke (QC), J1K 2R1 Canada

In the quest for novel magnetoelectric materials, we have grown, stabilized and explored the properties of La$_3$BAlMnO$_9$ (B = Co or Mn) thin films. In this paper, we report the influence of the growth parameters that promote B/Al/Mn ordering in the pseudo-cubic unit cell and their likely influence on the magnetic and multiferroic properties. The temperature dependence of the magnetization shows that La$_3$CoAlMnO$_9$ is ferromagnetic up to 190 K while La$_3$NiAlMnO$_9$ shows a T$_C$ of 130 K. The behavior of these films are compared and contrasted with related La$_2$BMnO$_6$ double perovskites. It is observed that the insertion of AlO$_6$ octahedra between CoO$_6$ and MnO$_6$ suppresses significantly the strength of the superexchange interaction, spin-phonon and spin-polar coupling.

# I. Introduction

Double perovskite oxides with chemical formula $A_2BMnO_6$ (*e.g.,* A = La or Bi, B= Co or Mn) have received considerable interest as magnetoelectric multiferroic materials [1-7]. These systems show a ferromagnetic insulator [8] behavior and a strong spin-phonon-polar coupling [1-6, 9-12]. In these systems, ferromagnetism with an insulator characteristic arises due to the 180° superexchange interaction between B-O-Mn cations, as described by Goodenough and Kanamori [8]. Unlike the disordered phase of $A_2BMnO_6$, B and Mn cations are alternatively stacked in the ordered phases [9]. Furthermore, the B ions (*i.e.,* Co or Ni) are in the 2+ oxidation states and the Mn ions are in the 4+ state while it is expected that in the disordered phase all these cations (Co/Mn/Ni) are in the 3+ oxidation states and/or in the mixed oxidation states [8]. In the self-ordered phase, superexchange interaction between $Co^{2+}$-O-$Mn^{4+}$ and $Ni^{2+}$-O-$Mn^{4+}$ leads to a ferromagnetic behavior reaching almost room temperature [6, 8]. In the case where the Co/Ni/Mn are disordered, an additional ferromagnetic transition around 140 K results from the $Co^{3+}$-O-$Mn^{3+}$ or $Ni^{3+}$-O-$Mn^{3+}$ superexchange paths and their magnetic dilution through $Co^{3+}$-O-$Co^{3+}$ or $Ni^{3+}$-O-$Ni^{3+}$ and $Mn^{3+}$-O-$Mn^{3+}$ bonds. This also illustrates that the magnetic properties of these oxides are a sensitive tool to characterize the presence of the cation ordering in these systems and have been successfully employed in various cases [2, 4-6].

Further in the ordered phase, these oxides possess charge ordering owing to the different oxidation states of B and Mn cations leading to a strong local electric field that couples with the magnetic order parameters. The magnetic field dependence of the dielectric constant of the ordered double perovskites (*e.g.,* $La_2CoMnO_6$ and $La_2NiMnO_6$) exhibits a large magnetodielectric (MD) effect in the vicinity of their ferromagnetic-to-paramagnetic transition [1, 3]. This illustrates the presence of a strong spin-polar coupling effect in these systems [1-3] and the possibility of their potential uses in real spintronic devices. Within the Landau Free energy framework, this MD effect[3] is explained by a $\gamma P^2 M^2$ coupling term, wherein $\gamma$ is the coupling constant, P is the electrical polarization, and M is the magnetization order parameters[6, 10, 11].

Despite the great deal of experimental results on the coupled properties of these oxides and their interpretation using phenomenological theory, very little is understood about the structural parameters that influences and controls the coupling mechanisms at microscopic level.[1-3, 6, 12-13]. Such studies are therefore crucial from both the technological and fundamental physics point of view. In this context, it therefore becomes interesting to explore closely related structures wherein one can meticulously modify and alter the local electric, phonon, and magnetic interactions and hence the coupling mechanism. In this perspective, we are also searching for unique routes to break the magnetic inversion symmetry which may lead to a ferroelectric ground state [14-18] by substituting a non-magnetic cation into the double perovskite unit cell while preserving the B/Mn cation ordering and therefore its ferromagnetic nature and polar behavior. Unlike the superlattices and composite methodology [19, 20], the beauty of this alternative approach is to design and to explore the magnetoelectric multiferroics by combining different building blocks in a single system while obtaining the polar ferromagnetic insulators.

To get further insights on the interplay of the coupling behavior and the magnetic properties, we therefore introduce additional complexity in these self-ordered double perovskites by inserting a non-magnetic $AlO_6$ octahedra in between the magnetic $BO_6$ and $MnO_6$ ones. This unique system may be viewed as the alternating blocks of $LaCoO_3$, $LaMnO_3$ and $LaAlO_3$ with three dimensional (3D) atomic-level ordering in which $LaAlO_3$ acts as a non-magnetic insulating spacer layer that separates the two magnetic layers, namely $LaCoO_3$ and $LaMnO_3$. The objective of such intricate structural modification is to alter the B-O-Mn superexchange interaction strength and eventually its dilution owing to the non-magnetic nature of Al, and hence the spin-polar coupling. Further, a polar layer separated by the insulating $LaAlO_3$ may also modify the nature of local electric field that arises owing to different oxidation states of B/Mn cations as said above. In contrast to this new compound, a superlattice [17] composed of one unit cell $La_2BMnO_6$ and one unit cell of $LaAlO_3$ will result only in 1D ordering (*i.e.,* in out-of-plane direction), while a 3D Co/Mn/Al ordering can be achieved in our system.

For this purpose, we doped 33.33% Al in equal parts at the B/Mn-sites in LNMO and LCMO. In the case of a random distribution, the resulting compounds become $LaNi_{0.33}Al_{0.33}Mn_{0.33}O_3$ and $LaCo_{0.33}Al_{0.33}Mn_{0.33}O_3$ respectively. Ideally, these systems once self-ordered would get the unit cell formulas $La_3NiAlMnO_9$ (LNAMO) and $La_3CoAlMnO_9$ (LCAMO) respectively [13]. It is also here important to note that the ionic radius of $Al^{3+}$ is very close to the value of Mn, Co, and/or Ni in an octahedral coordination. The LNAMO and LCAMO triple perovskite phases were stabilized in form of epitaxial films on $SrTiO_3$ (001) using a pulsed laser deposition (PLD) technique. We studied the structural and physical properties of these films using a variety of techniques. In this paper, we present the growth conditions that favour the self-ordering of Al/B/Mn cations in thin film form. The structural, magnetic, spin-phonon-polar coupling properties of these films are investigated. We compare and discuss the properties of these films with their double perovskite counter parts.

## II. Growth details

In our recent studies, we demonstrated that a stringent pulsed laser deposition conditions are required to obtain the ordered B/Mn configuration in the double perovskite phase. The detailed phase stability zone and growth boundary conditions for $La_2NiMnO_6$ (LNMO) and $La_2CoMnO_6$ (LCMO) films [4, 6] were mapped out using the magnetic, phonon, and/or structural properties of our films and available related literature. The presence of B/Mn cation ordering were illustrated in our samples by presence of a single magnetic transition around room temperature, emergence of additional Γ-phonon excitations in polarized Raman spectra, and/or the presence of superlattice lattice reflections in the X-ray diffraction and electron diffraction patterns of the films. The origins of these unique structural and physical properties are discussed in detail elsewhere.

The phase-stability zones to obtain the ordered double perovskites were used as a reference growth conditions to stabilize the LCAMO and LNAMO triple perovskite films. The epitaxial thin films of LCAMO and LNAMO on $SrTiO_3$ (001) were grown by ablating a stoichiometric target using a KrF excimer laser (λ = 248 nm) at 6Hz [13]. A growth temperature range of 720 to 860 °C in an $O_2$ pressure from 100 to 1000 mTorr was explored. The laser energy density on the target was about 1.1 J/cm$^2$ and target-to-substrate distance was about 5 cm. Following the ablation, the deposition chamber was filled with 400 Torr $O_2$ pressure and the samples were cooled down to room temperature at a rate of 10 °C/min. As-grown films have a thickness in the range of 50-123 nm, as measured by the stylus profilometer. Temperature dependence magnetic behavior and structural properties of as-grown triple perovskite films and the respective parental double perovskite films grown under the identical conditions [4, 6] were used to infer the parasitic-phase free LCAMO and LNAMO films and determine the growth conditions that promote a single phase formations. The stoichiometric polycrystalline LCAMO and LNAMO targets for ablation were synthesised by standard solid-state synthesis routes. The cationic stoichiometry of targets and films were examined and confirmed using energy dispersive spectroscopy associated with a scanning electron microscope. It shows that the compositional ratio of La: B: Mn: Al in the films was about 2:1:1:1. The error in cationic compositional analysis was about 10%. It is also important to note that these targets are multiphase and no trace of the triple perovskite structure can be identified (using x-ray diffraction and micro-Raman spectroscopy). Thus, these triple perovskite phases can only be stabilized in thin film form. Microstructural properties of our films and polycrystalline target will be published elsewhere.

## III. Results and Discussion

The epitaxial nature and the crystalline quality of the LCAMO and LNAMO films [13] were examined by X-ray diffractometer, operated in the θ-2θ and rocking curve modes using Cu-Kα radiation. XRD data were collected in the 2θ range of 15-60° in a step of 0.01°. XRD reflections of the $SrTiO_3$ substrate were also used as the internal standard to subtract the instrumental mechanical shift. A typical XRD pattern of a LCAMO film grown at 850 °C and under 600 mTorr is shown in Figure 1a. XRD patterns are indexed based on the pseudo-cubic perovskite unit cell notation. Fig.1a shows that the film is characterized by only one set of (*00l*) reflections (where *l* = 1, 2, 3, etc) demonstrating its preferred orientation. The out-of-plane lattice parameter, as computed from the XRD patterns, is about 4 Å, which is significantly larger than that of LCMO (~ 3.887Å). This suggests that the insertion of Al amid Co-O-Mn leads to the expansion of the pseudo-cubic LCMO unit cell leading to a minimization of the Coulombic energy arising from the polar character. Detailed XRD studies [13] on thin films of LNAMO also show a similar enhancement in its out-of-plane lattice parameter (3.995 Å) as compared to LNMO's value of 3.87 Å. To study further the coherent nature of the films, ω-scans (*i.e.*, rocking curves) were also measured on these (*00l*) reflections. A typical rocking curve measured on the (002) reflection of LCAMO film is shown in the inset of Fig. 1a. The full-width at half maximum (FWHM) of this curve is about 0.7° confirming that the films are grown coherently.

Our detailed study shows that films grown at low temperature and/or under low $O_2$ pressure comprised of multiple magnetic transitions and the ferromagnetic behavior akin to their parental double perovskites. Also it is important to note that such a growth conditions also do not favor a long range B/Mn ordering in double perovskites[4, 6, 13]. The shaded area in Fig. 1b presents the growth conditions that promote a B/Al/Mn ordering in this triple perovskite structure. As mentioned above, temperature dependence magnetic behavior and structural properties of as-grown triple perovskite films and the respective parental double perovskite films grown under the identical conditions were used to identify the growth parameters that promote to obtain the B/Al/Mn ordering and hence a parasitic phase free LCAMO and LNAMO films. For sake of comparison, we have also presented the phase-stability zone to obtain B/Mn ordering in their parental double perovskites. The shaded zone in this figure illustrates that a particular set of growth parameters are crucial to obtain either a Ni/Al/Mn or a Co/Al/Mn ordered phase. Unlike their double perovskite counterparts, only short-range ordered LNAMO and LCAMO films (see below) were obtained in these specific growth conditions. It therefore illustrates that a relatively higher growth temperature was essential to stabilize these triple perovskite films with respect to their double perovskite counterparts. It is here also important to note that the growth parameters shown in Fig. 1b represent the best possible conditions to obtain films with a minimum amount of defects. Our studies further suggest that a growth temperature above 860 °C (an accessibility limit to our PLD system) may be required to obtain long range ordered LCAMO triple perovskites.

The temperature dependence of the magnetization (M-T curves) under an applied magnetic field of 500 Oe and the magnetic field dependence of the magnetization (M-H loops) at 10K were measured using a superconducting quantum interference device (SQUID) magnetometer from Quantum Design. Typical M-T curves for LNMO, LNAMO, LCMO and LCAMO films are compared in Figure 2. The M-T curves for LCAMO and LNAMO reveal that the magnetization of these films are independent of temperature at low temperature confirming that these films are ferromagnetic in nature. The magnetic Curie temperatures, as estimated from M-T curves, are about 190 K and 130 K for LCAMO and LNAMO respectively. However, the value of LCMO FM-Tc is about 235 K and about 255 K for LNMO. This illustrates that the value of the FM transition temperatures of the triple perovskites are substantially lowered compared to the double perovskites.

The observation of a low value of FM-$T_c$ in the triple perovskites can be understood as follows: In a 180º-superexchange process [8], the magnetic transition is governed by the magnitude of the spin-transfer integral whose value is determined by the degree of orbital overlap in the B-O-Mn bonds. It ultimately depends on the bond lengths and its amplitude increases exponentially with the decreasing B-O-Mn bond lengths. The insertion of $AlO_6$ octahedra in LCMO/LNMO decreases the $MnO_6/BO_6$ overlap by provoking a new B-O-Al-O-Mn superexchange path. This ultimately reduces the overall strength of superexchange interaction between $MnO_6/BO_6$ and lowers the transition temperature. Nonetheless, this confirms that Al is doping the double perovskite unit cell. Additionally, B-O-Mn bonds may get diluted in antiferromagnetic/paramagnetic matrix arising from B-O-B and Mn-O-Mn bonds. Unlike LNAMO films, LCAMO films also display a secondary magnetic transition around 100 K, suggesting a partial amount of cationic disorder and short-range ordering.

To study further the ferromagnetic nature of LCAMO and LNAMO films, we measured the M-H loops at 10 K. Both oxides are characterized by well-defined M-H loops. LCAMO films display about 6.1 $\mu_B$/f.u. and a coercivity of about 4.5 kOe. On the other hand, LNAMO films display a 5.1 $\mu_B$/f.u. value of saturation magnetization and a coercivity of 0.17 kOe. Here f.u. stands for formula unit of ordered triple perovskite unit cell. For the calculation of the magnetization value of our films, 3.906 Å x 3.906 Å x 11.661Å for LCAMO and 3.906 Å x 3.906 Å x 11.998 Å for LNAMO were used as a formula unit volume. The values of saturation magnetization for triple perovskite oxides are found to be extremely close to their double perovskite counterparts [1-8]. Moreover, the magnetizations in these films are likely arising only from Co/Mn and Ni/Mn cations owing to the fact that Al is a non-magnetic cation. This illustrates that the films possess a large proportion of $Co^{2+}$ ($Ni^{2+}$) and $Mn^{4+}$ ions in the crystal structure while Al is likely to preserve its 3+ oxidation state. Thus, the presence of $Al^{3+}$ amid $B^{2+}$-O-$Mn^{4+}$ ions is expected to result in weaker spin-phonon-polar couplings due to the drastic reduction in the strength of the superexchange interactions.

Polarized Raman scattering is a powerful technique to study local structures and spin-phonon coupling in magnetic oxides [6, 10, 11]. It has been successfully used to demonstrate the presence of strong spin-phonon coupling in LNMO and LCMO films [6, 10, 11]. We used this technique to study the spin-phonon coupling behavior in LCAMO films. For this purpose, we measured the temperature dependence of the ~ 661 cm$^{-1}$ Raman active phonon frequency in the XX configuration (Inset of Fig. 3a) using a Labram 800 microscope spectrometer equipped with a He-Ne laser ($\lambda$ = 632.8 nm). The sample was cooled down in a Janis Research Supertran Cryostat and spectra were recorded using a nitrogen cooled charge coupled device (CCD) detector. Details about the polarization configurations used to collect the Raman spectra and their likely impact on the phonon excitations are reported elsewhere [6, 10, 11]. As one may notice, this particular phonon that corresponds to the stretching of the (Al/Co/Mn)$O_6$ octahedra starts to soften at the magnetic transition temperature revealing the first glimpse of a spin-phonon coupling similar to the one observed in the partially ordered and self-ordered LCMO [10, 11]. However, the net amplitude of softening in LCMO is about 10 cm$^{-1}$ whereas in LCAMO it is about 2.5 cm$^{-1}$. This suggests that LCAMO spin-phonon coupling is weak. In other words, the presence of Al$O_6$ octahedra amid Co$O_6$ and Mn$O_6$ octahedra significantly reduces the spin-phonon coupling strength as expected from the reduction of the superexchange interaction.

In order to study the spin-polar coupling in LCAMO, we measured the temperature dependence of the dielectric constant in LCAMO films grown on Nb:SrTiO$_3$

(001) under identical conditions [3, 13]. Dielectric measurements were carried out at 10 kHz using a 1V-AC excitation under different values of applied magnetic field (*viz.,* 0, 0.5, 10, 30 kOe) using a set-up described elsewhere[3]. Similarly to LCMO [3], LCAMO dielectric constant maximum is observed in the vicinity of its ferromagnetic transition temperature. Interestingly, a net change in the dielectric constant is observed under an applied magnetic field (Fig. 3b). As pointed out earlier in the self-ordered LCMO phase [3], this magnetodielectric response reaches a peak in the vicinity of the ferromagnetic-paramagnetic transition due to a $\gamma P^2 M^2$ term in the Landau free energy equation as a consequence of the coupling between the polar and magnetic order parameters [3, 12, 13]. The observation of a magnetodielectric effect in these films illustrates the presence of a spin-polar coupling in LCAMO. Despite these global similarities, the net amplitude of the magnetodielectric effect (~ 0.3%) in LCAMO is significantly smaller compared to LCMO's 3% amplitude increase around its ferromagnetic transition temperature[3]. This suggests that the spin-polar coupling effect has been suppressed in LCAMO and it indicates that the insertion of non-magnetic ions also reduces the spin-polar coupling strength of the double perovskites. Moreover, it smaller magnitude is consistent with the suppression of the spin-lattice coupling mentioned above due to the addition of the $AlO_6$ octahedra.

In summary, we have grown and investigated the structural, optical and magnetic properties of $La_3CoAlMnO_9$ and $La_3NiAlMnO_9$ triple perovskite films. Despite the presence of Al cation amid the B-O-Mn complex, these films display a ferromagnetic behavior. An evidence of spin-phonon-polar coupling is demonstrated. Our results may open alternative perspective to design and to control the coupling behavior in magnetoelectric perovskites.

We thank S. Pelletier and M. Castonguay for their technical support. This work was supported by *CIFAR*, *CFI*, *NSERC* (Canada), *FQRNT* (Québec) and the Université de Sherbrooke.

**Figure Captions:**

**Figure 1: (color online) (a)** A typical θ-2θ XRD pattern of the LCAMO film grown at 850 °C. The inset shows a rocking curve recorded around the (002) reflection. XRD peaks are indexed based on the pseudo-cubic notation. **(b)** Phase-stability diagram presenting the best conditions to grow quality films of short-range ordered $La_3BAlMnO_9$ and long-range ordered $La_2BMnO_6$. Arrows point out the respective phase-stability zones. The boundary of phase stability zone in fig.1b was determined by the magnetic, structural and physical properties of as-grown triple perovskite and parental double perovskites thin films [Refs 4 and 6].

**Figure 2: (color online)** Temperature dependence of the magnetization for **(a)** $La_3CoAlMnO_9$ and ordered $La_2CoMnO_6$ and **(b)** $La_3NiAlMnO_9$ and ordered $La_2NiMnO_6$ films. Respective insets present the magnetic field dependence of the magnetization for $La_3CoAlMnO_9$ and $La_3NiAlMnO_9$ measured at 10 K. Arrows show the magnetic transition points.

**Figure 3: (color online) (a)** temperature dependence of shift in the 661 $cm^{-1}$ mode revealing the presence of spin-phonon coupling in $La_3CoAlMnO_9$. Inset: Raman spectra measured at various temperatures for $La_3CoAlMnO_9$ films in the XX-configuration. **(b)** Temperature dependence of the dielectric constant of a $La_3CoAlMnO_9$ film measured under different magnetic fields.

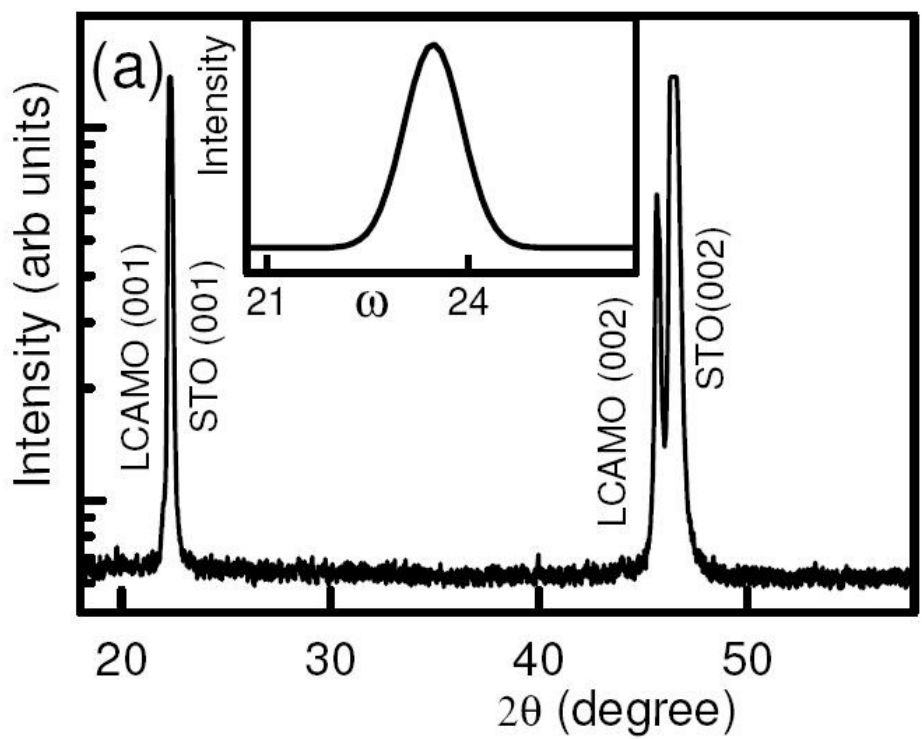
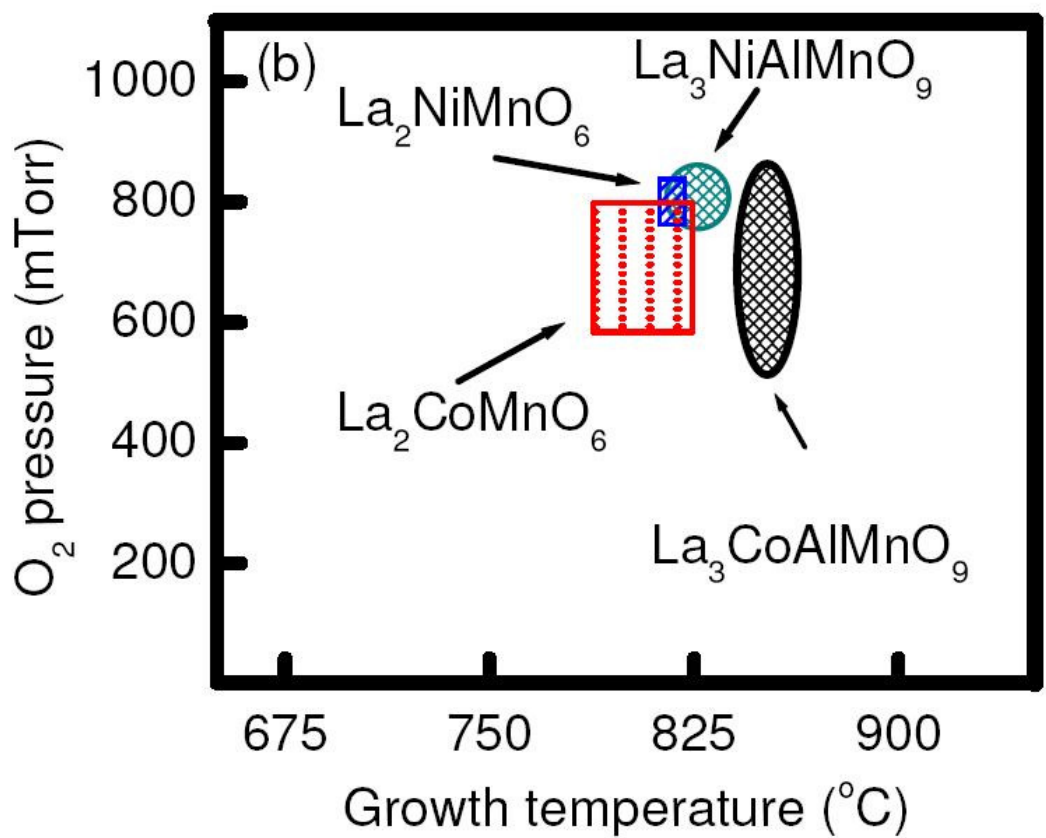

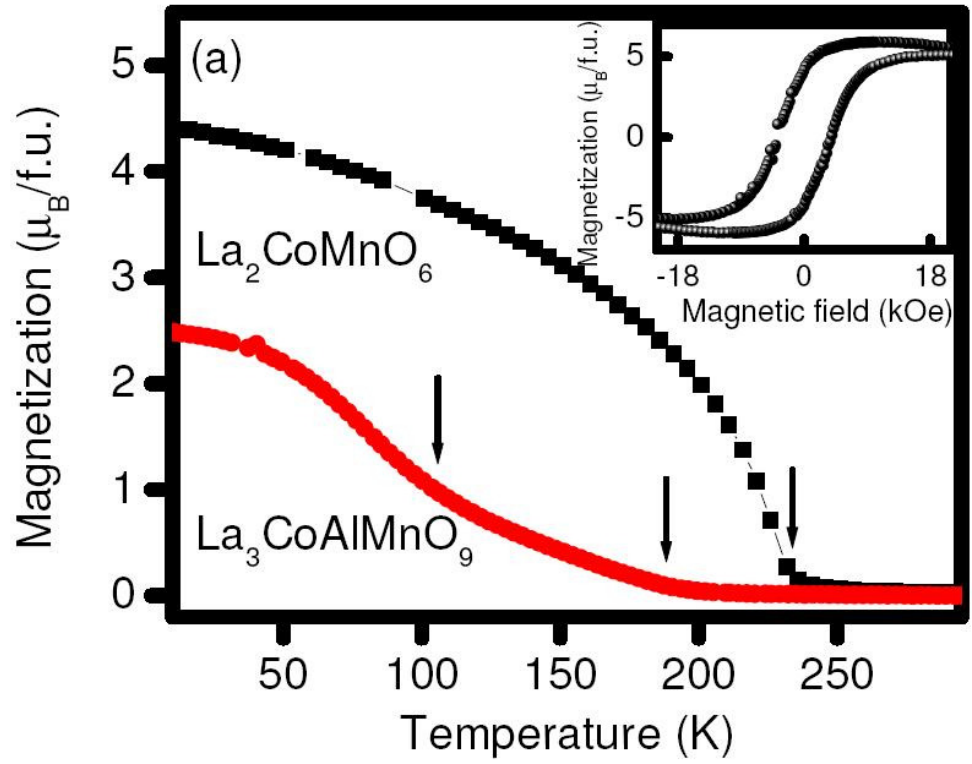
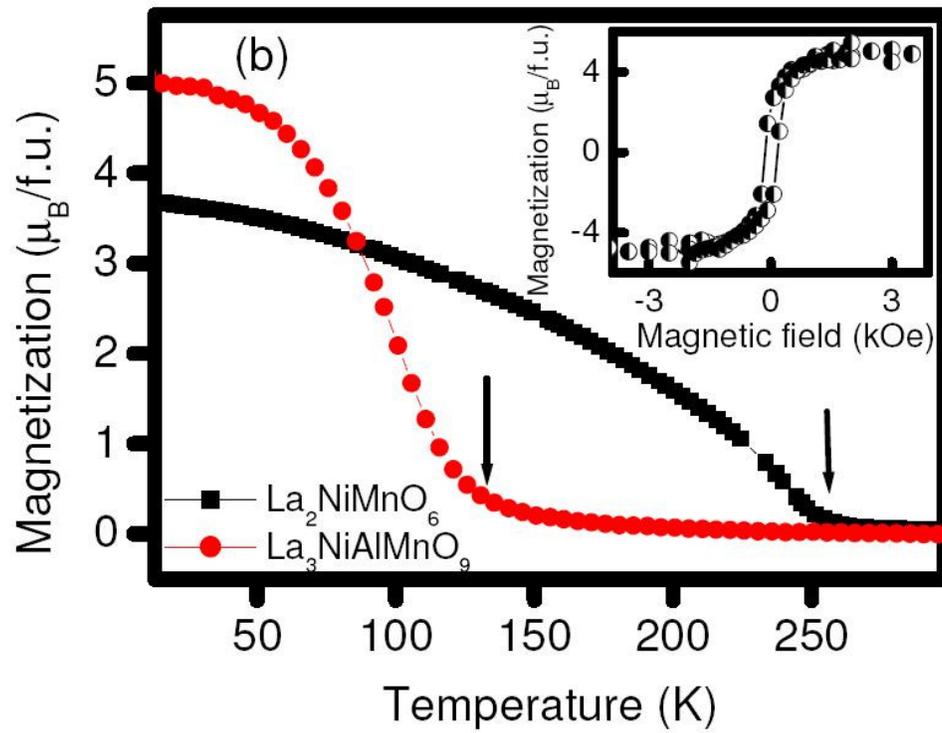

Fig. 2

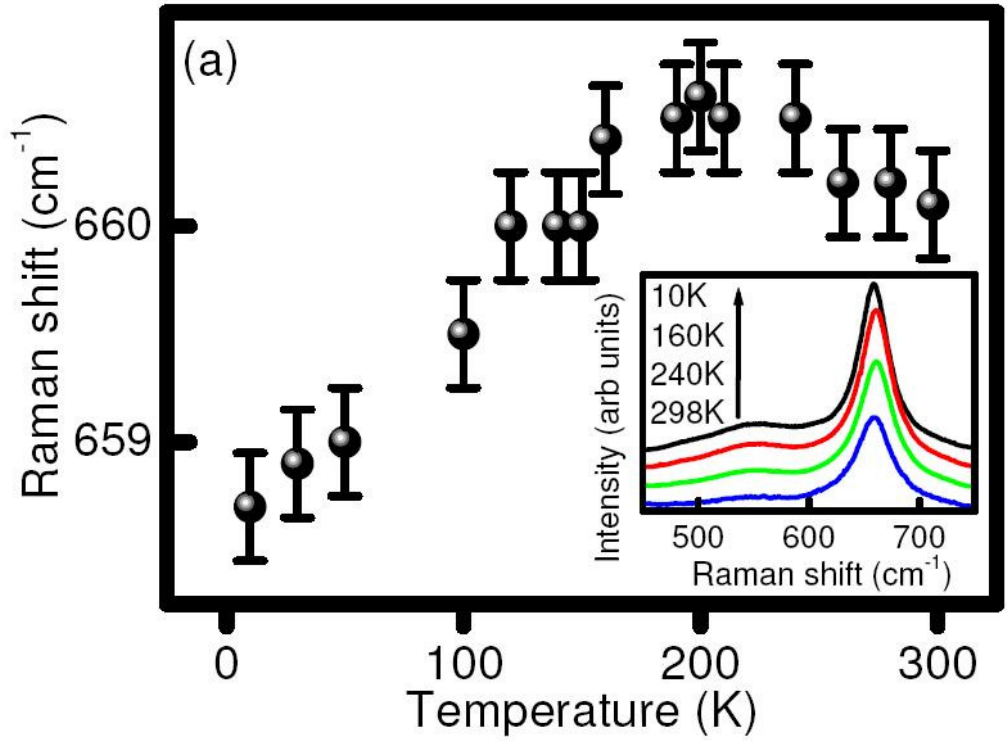

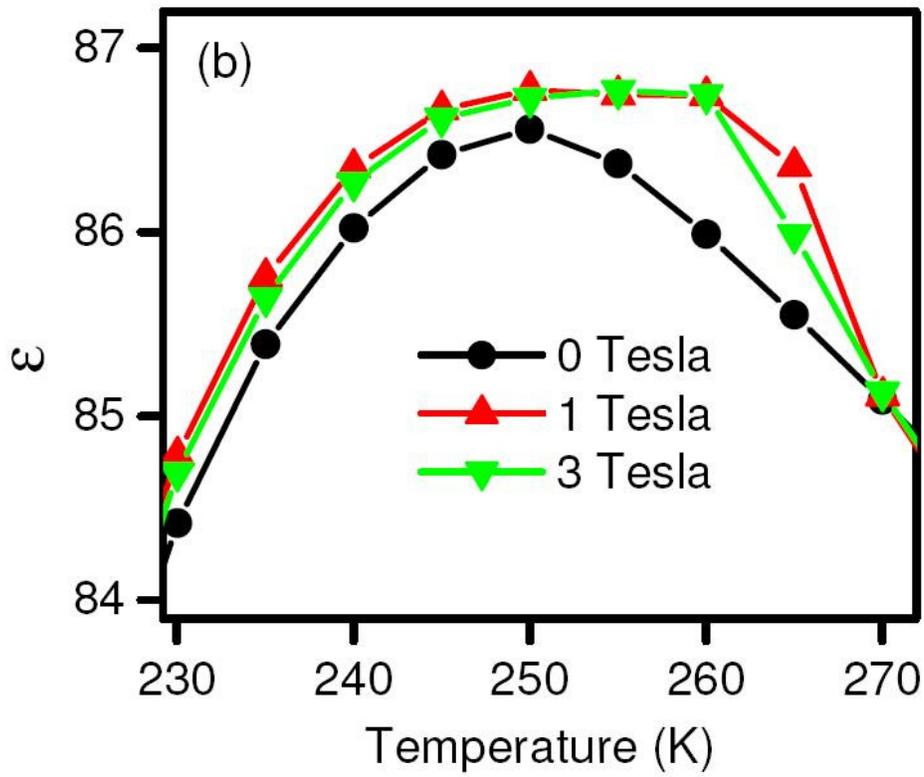

Fig. 3